\documentclass[copyright,creativecommons]{eptcs}

\providecommand{\event}{ACL2 2014} 
\usepackage{breakurl}             

\usepackage[fleqn]{amsmath}
\usepackage{amsthm}
\usepackage{amssymb}

\usepackage{graphicx}

\usepackage{listings}
\usepackage{txfonts}

\usepackage{url}
\newcommand{\keywordname}{Keywords:}
\newcommand{\keywords}[1]{\par\addvspace\baselineskip\noindent\keywordname\enspace\ignorespaces#1}

\begin{document}

\title{Formal Verification of Medina's Sequence of Polynomials for Approximating Arctangent}
\def\titlerunning{Polynomial Approximations to Arctangent in ACL2}
\def\authorrunning{Ruben Gamboa \& John Cowles}

\author{Ruben Gamboa 
\institute{University of Wyoming \\
Laramie, WY, USA}
\email{ruben@uwyo.edu}
\and
John Cowles
\institute{University of Wyoming \\
Laramie, WY, USA}
\email{cowles@uwyo.edu}
}

\maketitle

\begin{abstract}
The verification of many algorithms for calculating transcendental functions is based on polynomial approximations to these functions, often Taylor series approximations. However, computing and verifying approximations to the arctangent function are very challenging problems, in large part because the Taylor series converges very slowly to arctangent---a 57th-degree polynomial is needed to get three decimal places for $\arctan(0.95)$. Medina proposed a series of polynomials that approximate arctangent with far faster convergence---a 7th-degree polynomial is all that is needed to get three decimal places for $\arctan(0.95)$. We present in this paper a proof in ACL2(r) of the correctness and convergence rate of this sequence of polynomials. The proof is particularly beautiful, in that it uses many results from real analysis. Some of these necessary results were proven in prior work, but some were proven as part of this effort.

\keywords{Arctangent, taylor series, polynomial approximations.}
\end{abstract}

\section{Introduction}
\label{intro}

In this paper, we describe a formalization in ACL2(r) of a polynomial
approximation to arctangent. The obvious approach to approximating a
transcendental function is to use a general approximation scheme, such
as the Taylor Series. However, the Taylor Series for arctangent converges
very slowly:
\begin{equation}
\label{taylor-arctan}
\arctan(x) = x - \frac{x^3}{3} + \frac{x^5}{5} - \dots =
\sum_{k=0}^{\infty}{\frac{(-1)^k}{2k+1}x^{2k+1}}
\end{equation}
As Equation~\ref{taylor-arctan} shows, the denominators are growing at
the rate of $O(n)$, not $O(n!)$ as is the case for the Taylor series
of sine, cosine, or $e^x$. Consequently, the n$^\text{th}$ terms in
the series decrease much more slowly, and the convergence rate is
disastrous.

The long-term goal of this research project is to formally model the
x86 instructions that compute trigonometric, logarithmic, and
exponential functions~\cite{Rus:transcendentals}.  So it is of
practical importance to use a polynomial approximation that converges
more quickly to arctangent. A recent result of Medina's provides such
an approximation~\cite{Med:arctan}, and this paper describes a
formalization of that result in ACL2(r).

The paper is organized as follows.  In Section~\ref{arctan}, we
describe how the arctangent function can be introduced in
ACL2(r). Section~\ref{polys} presents a necessary detour into the
basic calculus of polynomials, including the rules for integrating and
differentiating polynomials. Section~\ref{medina} deals with Medina's
polynomial approximation. Finally, Section~\ref{conclusion} presents
some concluding remarks on the use of ACL2(r) for this project.

\section{The Arctangent in ACL2(r)}
\label{arctan}

\subsection{Introducing Arctangent}

We begin this discussion by introducing the arctangent function into
ACL2(r). From the perspective of ACL2(r), the exponential function
$e^x$ is the most  fundamental of the transcendental functions. It is
defined as a power series over the complex plane, and the
trigonometric functions sine and cosine are introduced in terms
of $e^x$. The tangent function itself is introduced as
the quotient of sine and cosine.

ACL2(r) allows the definition of inverse functions, such as
arctangent~\cite{GaCo:inverses}.  In order to introduce the inverse
function for $f(x)$, it is necessary to prove certain obligations
(which correspond to constraints in a hidden \texttt{encapsulate}):
\begin{itemize}
\item $f:D\rightarrow R$ is defined on interval $D$, and its range is
  the interval $R$.
\item $f$ is 1-to-1 over the domain $D$.
\item $f$ is continuous over $D$.
\item If $y\in R$, there are $x_1\in D$ and $x_2\in D$ such
  that $f(x_1) \le y \le f(x_2)$.
\end{itemize}
The challenge, then, is to prove that tangent has these properties, in
order to introduce its inverse, arctangent.

By convention, we chose the relevant domain of tangent to be $(-\pi/2,
\pi/2)$, and the range of tangent over this domain is the entire
number line $\mathbb{R}$. 

Next we show that tangent is 1-to-1 on the domain $(-\pi/2,
\pi/2)$. We do this with a little calculus. If we can show that the
derivative of tangent is positive on $(-\pi/2, \pi/2)$, then it must,
necessarily, be increasing over this range. Moreover, if tangent is
differentiable on $(-\pi/2, \pi/2)$, it must also be continuous on
that range. Thus, the derivative of tangent provides two of the needed
proof obligations.

Tangent is defined in ACL2(r) as $\tan(x) \equiv
\frac{\sin(x)}{\cos(x)},$ so its derivative follows from the product and
quotient
rules and the derivatives of sine and
cosine~\cite{GaCo:chain-rule,ReGa:automatic-differentiator}. the major
complication is proving that $\cos(x)$ is non-zero for $x\in (-\pi/2,
\pi/2)$. This was actually proven earlier, in part to define the 
constant $\pi$ in ACL2(r) as (twice) the first positive zero of
cosine~\cite{Gam:dissertation}!  It should be noted that the result of
this effort is that
\begin{align}
\frac{d(\frac{\sin(x)}{\cos(x)})}{dx} & = \frac{\sin(x)  [(-1)(-\sin(x))]}{\cos^2(x)} + \cos(x) \frac{1}{\cos(x)} \\
&= \frac{\sin^2(x)}{\cos^2(x)} + 1
\end{align}
It takes (proving and) using the trigonometric identity $\tan^2(x) + 1
= \sec^2(x)$ to reduce this expression to the familiar $\tan'(x) =
\sec^2(x)$. As mentioned previously, now that the derivative is known,
it follows directly that tangent is continuous on the desired interval.

To show that tangent is 1-to-1 on the interval, we use the fact that
the derivative $\sec^2(x)$ is positive on $(-\pi/2,
\pi/2)$. We found it surprising that it was not already proven in
ACL2(r) that a positive $f'$ guarantees increasing $f$. We
formalized this small result using the Mean Value Theorem
(MVT). If there are $x_1$ and $x_2$ such that $x_1 > x_2$
but $f(x_1) \le f(x_2)$, then by the MVT there is a point $c$ such that
$x_1 < c < x_2$ and $f'(c) = \frac{f(x_2)-f(x_1)}{x_2-x_1} \le 0$. Since
$f'$ is positive, no such point $c$ exists, hence no such $x_1$ and
$x_2$ can be found.

The final proof obligation is that for any $y\in\mathbb{R}$, we can
find $x_1$ and $x_2$ in $(-\pi/2, \pi/2)$ such that $\tan(x_1) \le y \le
\tan(x_2)$.  This turned out to be a significant challenge, which we
tackled in parts.

For the first part, suppose $0 \le y \le 1$. Then $\tan(0) \le y \le
\tan(\pi/4)$, since $\tan(0) = 0$, $\tan(\pi/4) = 1$, and tangent is 
an increasing function. So setting $x_1=0$ and $x_2=\pi/4$ will work.

Before tackling the second part, we find an important lower bound on
$\tan(x)$ whenever $\pi/4 \le x < \pi/2$. The lower bound is easily 
found since $\tan(x) = \sin(x)/\cos(x)$, sine is increasing on $[0,
\pi/2]$, and $\sin(\pi/4) = 1/\sqrt{2}$, so $\tan(y) \ge
1/(\sqrt{2}\cos(x))$ when $\pi/4 \le y < \pi/2$.

For the second part, suppose that $y>1$.
The lower bound on tangent above can be turned into a range on arctangent as
follows.  Since $y > 1$, it follows that $1/(\sqrt{2}y) \in (0,
1)$. In turn, this means that $\arccos(y) \in (0, \pi/2)$.  Actually,
since cosine is decreasing on $(0, \pi/2)$, and
$\cos(\pi/4)=1/\sqrt{2}$, $\arccos(y)$ is further restricted to $(\pi/4,
\pi/2)$.  So for $y > 1$, it follows that 
$\tan(0) \le y \le \tan(\arccos(1/(\sqrt{2}y)))$, so setting
$x_1=0$ and $x_2=\arccos(1/(\sqrt{2}y))$ will work.

The third and final part, when $y < 0$, can be derived from the results
above by observing that $\tan(-y) = -\tan(y)$, so it is sufficient to
find the bound for $\arctan(-y)$ and swap signs.

At this point, the proof obligations for inverse functions are
fulfilled, so we can introduce arctangent using \texttt{definv}.

\subsection{The Derivative of Arctangent}

The next step is to define the derivative of arctangent. The
derivative of inverse functions was proven in~\cite{GaCo:chain-rule}
and is given by 
\begin{equation}
\frac{d(f^{-1}(y))}{dy} = \frac{1}{f'(f^{-1}(y))}.
\end{equation}
This formula is valid only when $f'$ is never infinitesimally small in
the range of $y$.

In the previous section, we showed that the derivative of tangent is
$\sec^2(x) = 1/\cos^2(x)$.  This function achieves its minimum when
cosine achieves its maximum magnitude, i.e., when $\cos(x) = \pm
1$. Consequently, $\tan'(x) \ge 1$, so it is never infinitesimally
small. That means
\begin{equation}
\frac{d(\tan^{-1}(y))}{dy} = \frac{1}{\sec^2(\arctan(y))} =
\frac{1}{\tan^2(\arctan(y)) + 1} = \frac{1}{y^2+1}.
\end{equation}

The Fundamental Theorem of Calculus (FTC) was first proved in ACL2(r)
in~\cite{Kau:ftc}, and we recently redid that proof to make the final
statement of the FTC more direct. Using this result, it follows that
\begin{equation}
\int_{a}^{b}{\frac{dx}{1+x^2}} = \arctan(b) - \arctan(a).
\end{equation}
This result will play a key role in Section~\ref{medina}.

\section{Polynomial Calculus}
\label{polys}

\subsection{The Derivative and Integral of $x^n$}

We now turn our attention to the derivative and integral of the
function $x^n$. Because this is really a binary function, of both
$x$ and $n$, it illustrates the difficulties of working with the
non-standard definition of derivative.  For example, a direct way of
proving that $\frac{d(x^n)}{dx} = n \cdot x^{n-1}$ is by using
induction, invoking the product rule during the inductive step. The
problem is that the non-standard definition of differentiability
requires that, $small(\epsilon) \Rightarrow \frac{(x+\epsilon)^n -
  x^n}{\epsilon} \approx n \cdot x^{n-1}$. This is a non-classical
formula, so it cannot be proved using functional instantiation with a
pseudo-lambda expression, e.g., $f(x) \rightarrow (\lambda (x)
x^n)$.

That is part of the motivation behind proving in ACL2(r) that the
$\epsilon$-$\delta$ definition of derivative is equivalent to the
non-standard definition used in ACL2(r)~\cite{CoGa:equivalences}.  
Indeed, using the $\epsilon$-$\delta$ definition of derivative, it is
possible to prove the derivative of $x^n$ by induction. However, there
are still potential pitfalls. In particular, the key lemma in the
inductive step requires the use of the product rule, $(f \times g)' =
f \times g' + f' \times g$.  But the proof obligations of the
functional instantiation include the theorem $\frac{d(x^{n-1})}{dx} =
(n-1) \cdot x^{n-2}$. This is part of the induction hypothesis, but
injecting hypotheses into proof obligations of functional
instantiation is a difficult problem.

So we opted for a slightly more general approach.  There are two
different ways of writing $x^n$ in ACL2(r):
\begin{itemize}
\item \texttt(expt x n)
\item \texttt(raise x n)
\end{itemize}
The \texttt{expt} function is identical to its counterpart in ACL2, so
it is defined by induction on \texttt{n} (which must be an integer, not
necessarily a natural number). The \texttt{raise} function is defined
using $x^n = e^{n \ln(x)}$. For integer exponents $n$, these two
definitions are known to be equal.

The idea, then, is to use the derivative of $e^{n \ln(x)}$ to find
the derivative of $x^n$. Previously, we had shown that the derivative
of $e^x$ is precisely $e^x$~\cite{ReGa:automatic-differentiator}. With
the use of the Chain Rule~\cite{GaCo:chain-rule} and the derivative of
$\ln(x)$~\cite{ReGa:automatic-differentiator}, this means that
\begin{align}
\frac{d(x^n)}{dx} &= \frac{d(e^{n\ln(x)})}{dx} \\
                          &= n \frac{1}{x} e^{n\ln(x)} \\
                                     &= n \frac{1}{x} x^n \\
                                     &= n x^{n-1}.
\end{align}
However, this derivation makes several hidden assumptions that need to
be addressed.

The first problem is that the derivative of $\ln(x)$ is only known for
$x>0$.  (While the function $\ln(x)$ is defined for all non-zero
complex numbers, derivatives in ACL2(r) are restricted to real-valued
functions of real numbers.) So for positive values of $x$, this
argument does hold, and we proved that
\begin{equation}
x>0 \Rightarrow \frac{d(x^n)}{dx} = n x^{n-1}.
\end{equation}

When $x<0$, $e^{n\ln(x)}$ isn't even necessarily defined over the
reals, e.g., $(-1)^{\frac{1}{2}} = e^{\frac{1}{2} \ln(-1)} = i \not\in
\mathbb{R}$. However, we can restrict $n$ to range over the integers,
and then $x^n$ is defined even for negative $n$. Our approach was to
show that whenever $x<0$, 
\begin{align}
x^n &= e^{n \ln(x)} \\
       &= e^{n \ln(-|x|)} \\
       &= e^{n (\ln(|x|) + i \pi)} \\
       &= e^{n \ln(|x|) + i \pi n} \\
       &= e^{n \ln(|x|)} e^{i \pi n} \\
       &= e^{n \ln(|x|)} (-1)^n
\end{align}
In the last step, $(-1)^n$ can be represented using either
\texttt{raise} or \texttt{expt}, since $n$ is restricted to the
integers. This means that $(-1)^n$ is equal to $1$ when $n$ is even
and $-1$ when $n$ is odd, and these cases can be considered
separately. At this point, the derivative of $x^n$ can be reduced to
the case where $x>0$, since $|x|>0$. This shows that 
\begin{equation}
x<0 \wedge n \in \mathbb{Z} \Rightarrow \frac{d(x^n)}{dx} = n x^{n-1}.
\end{equation}

That leaves the case when $x=0$. Again, we restrict ourselves to the
case of integer $n$, because it is possible for $\epsilon$ to be
infinitessimally close to $0$ yet still be negative. Moreover, $n$
cannot be negative, because in that case $0^n$ is undefined.
When $n=0$, $x^n=1$, so the derivative of $x^n$ is 0, which is equal
to $n x^{n-1} = 0\cdot x^{-1} = 0$. Note: This uses the fact that $1/0 =
0$ according to the axioms of ACL2.
When $n>0$, $0^n = 0$ and $|\epsilon^n| \le
|\epsilon|$ for $|\epsilon| < 1$. If $n=1$, then
$\epsilon^n=\epsilon$, and the derivative of $x^n$ is just 1, and
since $0^0=1$, this is exactly the same as $n x^{n-1} = 1 \cdot 0^{0}
= 1$.  When $n>1$, for infinitesimal $\epsilon$, $\epsilon^n \approx 0
= n 0^{n-1} = n \cdot 0$. So we have shown that 
\begin{equation}
x=0 \wedge n \in \mathbb{N} \Rightarrow \frac{d(x^n)}{dx} = n x^{n-1}.
\end{equation}

Combining these results, we have that
\begin{equation}
\label{deriv-expt}
\left[
(x>0) \vee 
(x<0 \wedge n \in \mathbb{Z}) \vee
(x=0 \wedge n \in \mathbb{N}) \right] \Rightarrow \frac{d(x^n)}{dx} = n x^{n-1}.
\end{equation}
It is interesting that so many hypotheses are needed for this result,
which is taken for granted in calculus. However, the assumption there
is that the result holds only when all expressions in the theorem are
defined. This is a powerful assumption that hides hypotheses.

Before proceeding, we would like to make the following
observation. Many of the theorems require hypotheses such as $n \in
\mathbb{Z}$. Since $n$ is not one of the parameters of the function
$f$ that is being functionally instantiated, these arguments have to
be ``infected'' when using functional instantiation. One of the
traditional approaches is to use a pseudo-lambda term with a condition
and a default value, as in the following:
\begin{lstlisting}
:functional-instance useful-theorem
                     (f (lambda (x)
                         (if (not (integerp n))
                             0
                           (expt x n))))
\end{lstlisting}
However, since many such functions need to be instantiated, it is not
always obvious how to define the ``unintended domain'' cases so that
the constraints of all the combined functions hold. So we found it more
productive to move these hypotheses into the definitions, as in the
following:
\begin{lstlisting}
(defun raise-to-int (x n)
  (raise (realfix x) (ifix n)))
\end{lstlisting}
Then we proved the required theorems about the ``fixed'' functions,
and only later raised the hypotheses to the statements as in Equation~\ref{deriv-expt}.

Once the derivative of $x^n$ is known, it is a simple matter to invoke
the FTC to find the integral of $x^n$:
\begin{equation}
\left[
(x>0) \vee 
(x<0 \wedge n \in \mathbb{Z}) \vee
(x=0 \wedge n \in \mathbb{N}) \right] \Rightarrow
\int_{a}^{b}{x^n dx} = \frac{b^{n+1}}{n+1} - \frac{a^{n+1}}{n+1}.
\end{equation}

\subsection{The Derivative and Integral of Polynomials}

It is now time to extend the results in the previous section to
polynomials. The first challenge is to capture the notion of
polynomials in ACL2(r), and we chose to use the characterization 
described in~\cite{GaCo:cantor-trio}. Polynomials are encoded as lists
of coefficients, with the first coefficient being the constant term,
and subsequent coefficients corresponding to higher powers of $x$. For
example, the polynomial $3 + x^2$ is encoded as the list \texttt{(3 0 1)}.
The function \texttt{eval-polynomial} evaluates a polynomial at a point, and
what we have to show is that its derivative is also a polynomial.
That particular function used the following recursive scheme:
\begin{equation}
evalpoly(cons(c,rest), x) = c + x\cdot evalpoly(rest, x)
\end{equation}
It is an easy challenge to define an alternative execution based on a
scheme that uses $x^n$:
\begin{equation}
evalpoly(cons(c,rest), x, n) = c\cdot x^n + evalpoly(rest, x, n+1)
\end{equation}
Once these two functions are proved equivalent, the results from the
previous section can be used directly.

So the first step is to define the list of coefficients of the
derivative of a polynomial. This is easily done, e.g., as in the
following definition:
\begin{lstlisting}
(defun derivative-polynomial-aux (poly n)
  (if (and (real-polynomial-p poly)
           (natp n)
           (consp poly))
      (if (< 0 n)
          (cons (* n (car poly))
                (derivative-polynomial-aux (cdr poly) (1+ n)))
        (derivative-polynomial-aux (cdr poly) (1+ n)))
    nil))
\end{lstlisting}
The proof that this polynomial is the derivative of the original
polynomial can proceed by induction. Recall that one of the
complications described in the previous section is the difficulty of
pushing the inductive hypothesis into the proof obligations of a
functional instantiation. However, the key lemma that is required in
this case is that $(f+g)'(x) = f'(x)+g'(x)$. The proof of this lemma
is easy enough that it can be carried out as part of the
induction. The trick is to do induction such that $\langle poly, n,
\epsilon \rangle \rightarrow \langle cdr(poly), n+1, \epsilon/2
\rangle$.

As before, once the derivative of polynomials is established, it is 
easy to invoke the FTC in order to introduce the integral of
polynomials. We defined a function similar to
\texttt{derivative-polynomial-aux} that computes the coefficient of
the integral.

\section{Medina's Result}
\label{medina}

Now that all preliminaries have been dealt with, we can formalize
Medina's main result. In order to make arctangent more tractable,
Medina first reduces the domain of arctangent to $[0,1]$. He can do
this by using the following lemmas:
\begin{gather}
x > 1 \Rightarrow \arctan(x) = \frac{\pi}{2} - \arctan\left(\frac{1}{x}\right) \label{medina-lemma-1}\\
x < 0 \Rightarrow \arctan(x) = -\arctan(-x) \label{medina-lemma-2}
\end{gather}
The proof of Equation~\ref{medina-lemma-1} follows by proving that the
tangent of both sides is equal, and then using the uniqueness of
inverse functions (in the appropriate
domain). Equation~\ref{medina-lemma-2} follows even more directly
using the same approach. Incidentally, neither of these lemmas
requires the given hypothesis.

Now that these lemmas are proved, we can restrict $x$ to the range
$x\in[0,1]$. Medina defines the following sequence of polynomials:
\begin{align}
p_1(x) &= 4 - 4x^2 + 5x^4 - 4x^5 + x^6 \\
p_m(x) &= x^4(1-x)^4 p_{m-1}(x) + (-4)^{m-1} p_1(x)
\end{align}
The first step is to find a more direct way of writing $p_m$.  For
$m\ge 2$, the polynomial can be written as follows:
\begin{equation}
\label{medina-alternative}
p_m(x) = \frac{x^{4m}(1-x)^{4m} + (-4)^m}{1+x^2}.
\end{equation}
This is not obviously a polynomial, but $1+x^2$ is actually a factor
of the numerator. But since the structure is not clearly that of a
polynomial, we introduced the functions $p_m$ explicitly, instead of
using \texttt{eval-polynomial}.

The proof of Equation~\ref{medina-alternative} is quite involved,
although it requires only induction on $m$ and elementary algebra.
The difficulty comes from the necessary algebraic manipulations.

We next focus on the term $x(1-x)=x-x^2$ when $x\in[0,1]$. The
derivative of this polynomial is $1-2x$, and this is zero when
$x=1/2$. In prior work, we had proved the Extreme Value Theorem that
says the derivative is zero when the function achieves a maximum or
minimum~\cite{Gam:dissertation}. Unfortunately, that is not the lemma
that is required here. Instead, what is needed is to show that when
the derivative is zero \emph{and some other conditions hold}, the
function is at a maximum.  The ``other conditions'' can vary, but we
chose to formalize the First-Derivative Test. That
is, if the derivative is positive for all $x<a$, zero at $a$, and
negative for all $x>a$, then $f$ achieves a maximum at $a$.
More precisely, the variable $x$ is restricted to range over some
interval $I$ containing $a$, not over all reals---although in this
case, that would have been sufficient. Since $x(1-x)$ achieves a
maximum at $1/2$, we have that $x(1-x) \le 1/4$ for all $x\in[0,1]$.
Moreover, since $x(1-x)\ge 0$ when $x\in[0,1]$, it follows that
\begin{equation}
x^{4m}(1-x)^{4m}\le \left(\frac{1}{4}\right)^{4m}.
\end{equation}
Now, $1+x^2 \ge 1$, so we have also shown that 
\begin{equation}
\frac{x^{4m}(1-x)^{4m}}{1+x^2} \le \left(\frac{1}{4}\right)^{4m}.
\end{equation}
Taking the integral of both sides shows the following:
\begin{align}
\int_{0}^{x}\frac{t^{4m}(1-t)^{4m}}{1+t^2} dt &\le \int_{0}^{x} \left(\frac{1}{4}\right)^{4m} dt \\
&=\left(\frac{1}{4}\right)^{4m} x \\
&\le \left(\frac{1}{4}\right)^{4m} \label{medina-bound}
\end{align}
Note that the last step follows only because $x\in[0,1]$.

We now return to Equation~\ref{medina-alternative}, which we reproduce
below:
\begin{equation}
p_m(x) = \frac{x^{4m}(1-x)^{4m} + (-4)^m}{1+x^2}.
\end{equation}
This can be rewritten as follows:
\begin{align}
\frac{x^{4m}(1-x)^{4m}}{1+x^2}  &= p_m(x) + \frac{(-4)^m}{1+x^2} \\ 
                                                &= p_m(x) - \frac{(-1)^{m+1}4^m}{1+x^2}.
\end{align}
Notice that the left-hand side is non-negative for $x\in [0,1]$, so
the right-hand side must be non-negative as well. We will use that
observation in the next step, but first we take integrals of both
sides and use Inequality~\ref{medina-bound}:
\begin{align}
p_m(t) - \frac{(-1)^{m+1}4^m}{1+t^2} &= \frac{t^{4m}(1-t)^{4m}}{1+t^2} \\
\int_{0}^{x} p_m(t) - \frac{(-1)^{m+1}4^m}{1+t^2} dt &= \int_{0}^{x} \frac{t^{4m}(1-t)^{4m}}{1+t^2} dt 
                                                                                \le \left(\frac{1}{4}\right)^{4m}
\end{align}
The next step is to divide the last equation by $(-1)^{m+1}4^m$.  This
can change the direction of the inequality, but since both terms are
positive (as discussed above), the magnitude of absolute values is
preserved.  This results in the following:
\begin{equation}
\left| \int_{0}^{x} \frac{p_m(t)}{(-1)^{m+1}4^m} - \frac{1}{1+t^2} dt \right| \le \left|\frac{1}{4}\right|^{5m}
\end{equation}

Now, we use the derivative of arctangent to integrate the second term
in the integral.
\begin{align}
\left| \int_{0}^{x} \frac{p_m(t)}{(-1)^{m+1}4^m} - \frac{1}{1+t^2} dt \right| &\le \left|\frac{1}{4}\right|^{5m} \\
\left| \int_{0}^{x} \frac{p_m(t)}{(-1)^{m+1}4^m} dt - \int_{0}^{x}\frac{1}{1+t^2} dt \right| &\le \left|\frac{1}{4}\right|^{5m} \\
\left| \int_{0}^{x} \frac{p_m(t)}{(-1)^{m+1}4^m} dt - \arctan(x) \right| &\le \left|\frac{1}{4}\right|^{5m} 
\end{align}
All that is left is to define the polynomial approximation:
\begin{equation}
h_m(x) \equiv \int_{0}^{x} \frac{p_m(t)}{(-1)^{m+1}4^m} dt.
\end{equation}
The previous results show that $h_m(x)$ is a good approximation to
arctangent.  In particular,
\begin{equation}
\left| h_m(x)  - \arctan(x) \right| \le \left|\frac{1}{4}\right|^{5m}.
\end{equation}
The $1/4^{5m}$ term on the right-hand side shows that the convergence
is quite good.

As before, it is not at all obvious that $h_m(x)$ is actually a
polynomial. But this does follow because $p_m(x)$ is a polynomial,
the other term inside the integral is a constant, and the integral of
a polynomial is also a polynomial. It would be more satisfying,
however, to have an expression for $h_m(x)$ that is an actual list of
coefficients. Medina does derive a closed form for $p_m$, and hence
for $h_m$, and we have formalized that proof in ACL2(r). The details
of that proof involve mostly tedious algebra, so we do not present
them here.

\section{Conclusion}
\label{conclusion}

This paper formalized a result of Medina's which defined a polynomial
approximation to arctangent that converges quickly. The proof made
heavy use of results from prior work formalizing real analysis, such
as the FTC, the MVT, composition rules for derivatives, etc. In
addition, a handful of results were missing and were proved as part of
this effort, such as the First Derivative Test.

In some ways, the result is an obvious candidate for ACL2(r), as
opposed to ACL2, since the final theorem uses the transcendental
function arctangent:
\begin{equation}
\label{medina-eqn-1}
\left| h_m(x)  - \arctan(x) \right| \le \left|\frac{1}{4}\right|^{5m}.
\end{equation}
However, one can envision a way of proving this result in ACL2, and
this is not unreasonable, since ACL2 has been used in the past to
prove the correctness of hardware approximations of functions that do
not technically exist in ACL2, such as the square root function. The
key step is to start with an approximation of the given function, and
then show that some other (e.g., faster) approximation is also close.

For instance, instead of using arctangent, we could start with the
Taylor approximation in Equation~\ref{taylor-arctan}. In particular,
the polynomial $T_n(x)$ could be defined as the Taylor approximation
of order $n$. This could lead to a theorem such as the following:
\begin{equation}
\label{medina-taylor-eqn-1}
\left| h_m(x)  - T_n(x) \right| \le \left|\frac{1}{4}\right|^{5m}.
\end{equation}
The problem is that it is not obvious how to compare $h_m$ and
$T_n$. Certainly, the theorem will not hold when $n=m$.  After all,
$h_m$ should converge to arctangent much more quickly than $T_n$!
Moreover, a recent discussion in the ACL2 mailing list has brought
attention to the fact that proving that two different series converge
to the same value can be very difficult in ACL2. The solution
suggested by the experts in the mailing list is to show that each of
the two series converges to some function, and that the functions the
series converge to are the same. But such a strategy could not be
carried out in this case, since arctangent is provably not in ACL2.
E.g., $\arctan(1) = \pi/4$ is not a number in ACL2, since it is
irrational.

So we believe that it is necessary to have support for the reals in
order to reason about results such as Inequality~\ref{medina-eqn-1} and
even Inequality~\ref{medina-taylor-eqn-1}, and we are delighted that
enough of real analysis has been formalized in ACL2(r) that the
formalization effort was mostly focused on the results specific to the
problem at hand, and (with the exception of the First Derivative Test)
not on more fundamental results.

\bibliographystyle{eptcs}
\bibliography{rag}

\providecommand{\JnodotS}{J S}
\begin{thebibliography}{1}
\providecommand{\bibitemdeclare}[2]{}
\providecommand{\surnamestart}{}
\providecommand{\surnameend}{}
\providecommand{\urlprefix}{Available at }
\providecommand{\url}[1]{\texttt{#1}}
\providecommand{\href}[2]{\texttt{#2}}
\providecommand{\urlalt}[2]{\href{#1}{#2}}
\providecommand{\doi}[1]{doi:\urlalt{http://dx.doi.org/#1}{#1}}
\providecommand{\bibinfo}[2]{#2}

\bibitemdeclare{misc}{CoGa:equivalences}
\bibitem{CoGa:equivalences}
\bibinfo{author}{John \surnamestart Cowles\surnameend} \&
  \bibinfo{author}{Ruben \surnamestart Gamboa\surnameend}
  (\bibinfo{year}{2014}): \emph{\bibinfo{title}{Equivalence of the Traditional
  and Non-Standard Definitions of Concepts from Real Analysis}}.
\newblock \bibinfo{howpublished}{Under review}.

\bibitemdeclare{phdthesis}{Gam:dissertation}
\bibitem{Gam:dissertation}
\bibinfo{author}{R.~\surnamestart Gamboa\surnameend} (\bibinfo{year}{1999}):
  \emph{\bibinfo{title}{Mechanically Verifying Real-Valued Algorithms in
  ACL2}}.
\newblock Ph.D. thesis, \bibinfo{school}{The University of Texas at Austin}.

\bibitemdeclare{inproceedings}{GaCo:chain-rule}
\bibitem{GaCo:chain-rule}
\bibinfo{author}{R.~\surnamestart Gamboa\surnameend} \&
  \bibinfo{author}{J.~\surnamestart Cowles\surnameend} (\bibinfo{year}{2009}):
  \emph{\bibinfo{title}{The Chain Rule and Friends in {ACL2(r)}}}.
\newblock In: {\sl \bibinfo{booktitle}{Proceedings of the Eighth International
  Workshop of the ACL2 Theorem Prover and its Applications (ACL2-2009)}}.

\bibitemdeclare{inproceedings}{GaCo:inverses}
\bibitem{GaCo:inverses}
\bibinfo{author}{R.~\surnamestart Gamboa\surnameend} \&
  \bibinfo{author}{J.~\surnamestart Cowles\surnameend} (\bibinfo{year}{2009}):
  \emph{\bibinfo{title}{Inverse Functions in {ACL2(r)}}}.
\newblock In: {\sl \bibinfo{booktitle}{Proceedings of the Eighth International
  Workshop of the ACL2 Theorem Prover and its Applications (ACL2-2009)}}.

\bibitemdeclare{inproceedings}{GaCo:cantor-trio}
\bibitem{GaCo:cantor-trio}
\bibinfo{author}{R.~\surnamestart Gamboa\surnameend} \&
  \bibinfo{author}{J.~\surnamestart Cowles\surnameend} (\bibinfo{year}{2012}):
  \emph{\bibinfo{title}{A {C}antor Trio: Denumerability, the Reals, and the
  Real Algebraic Numbers}}.
\newblock In: {\sl \bibinfo{booktitle}{Proc of the Third Conference on
  Interactive Theorem Proving (ITP-2012)}}, \doi{10.1007/978-3-642-32347-8_5}.

\bibitemdeclare{incollection}{Kau:ftc}
\bibitem{Kau:ftc}
\bibinfo{author}{M.~\surnamestart Kaufmann\surnameend} (\bibinfo{year}{2000}):
  \emph{\bibinfo{title}{Modular Proof: The Fundamental Theorem of Calculus}}.
\newblock In \bibinfo{editor}{M.~\surnamestart Kaufmann\surnameend},
  \bibinfo{editor}{P.~\surnamestart Manolios\surnameend} \&
  \bibinfo{editor}{{\JnodotS}.~\surnamestart Moore\surnameend}, editors: {\sl
  \bibinfo{booktitle}{Computer-Aided Reasoning: ACL2 Case Studies}},
  chapter~\bibinfo{chapter}{6}, \bibinfo{publisher}{Kluwer Academic Press},
  \doi{10.1007/978-1-4615-4449-4}.

\bibitemdeclare{article}{Med:arctan}
\bibitem{Med:arctan}
\bibinfo{author}{Herbert \surnamestart Medina\surnameend}
  (\bibinfo{year}{2006}): \emph{\bibinfo{title}{A Sequence of Polynomials for
  Approximating Inverse Tangent}}.
\newblock {\sl \bibinfo{journal}{American Mathematical Monthly}}
  \bibinfo{volume}{113}(\bibinfo{number}{2}), pp. \bibinfo{pages}{156--161},
  \doi{10.2307/27641866}.

\bibitemdeclare{inproceedings}{ReGa:automatic-differentiator}
\bibitem{ReGa:automatic-differentiator}
\bibinfo{author}{P.~\surnamestart Reid\surnameend} \&
  \bibinfo{author}{R.~\surnamestart Gamboa\surnameend} (\bibinfo{year}{2011}):
  \emph{\bibinfo{title}{Automatic Differentiation in ACL2}}.
\newblock In: {\sl \bibinfo{booktitle}{Proc of the Second Conference on
  Interactive Theorem Proving (ITP-2011)}}, \doi{10.1007/978-3-642-22863-6_23}.

\bibitemdeclare{misc}{Rus:transcendentals}
\bibitem{Rus:transcendentals}
\bibinfo{author}{David \surnamestart Russinoff\surnameend}
  (\bibinfo{year}{2007}): \emph{\bibinfo{title}{Modeling the Transcendental
  Instructions with Elementary Polynomial Approximations}}.
\newblock
  \bibinfo{howpublished}{\url{http://www.russinoff.com/papers/transcendentals.pdf}}.

\end{thebibliography}

\end{document}